\documentclass[apj]{emulateapj}
\shorttitle{M31 Transverse Velocity and Local Group Mass}
\shortauthors{van der Marel \& Guhathakurta}

\newcommand{\etal}{{et al.~}}
\newcommand{\lta}{\lesssim}

\newcommand{\kms}{\>{\rm km}\,{\rm s}^{-1}}

\newcommand{\uasyr}{\>\mu{\rm as}\,{\rm yr}^{-1}}
\newcommand{\Gyr}{\>{\rm Gyr}}
\newcommand{\kpc}{\>{\rm kpc}}
\newcommand{\Mpc}{\>{\rm Mpc}}
\newcommand{\Msun}{\>{\rm M_{\odot}}}

\begin{document}

\title{M31 Transverse Velocity and Local Group Mass from Satellite Kinematics}

\author{Roeland P.~van der Marel}
\affil{Space Telescope Science Institute, 3700 San Martin Drive, 
       Baltimore, MD 21218}

\author{Puragra Guhathakurta}
\affil{UCO/Lick Observatory, Department of Astronomy and Astrophysics, 
       University of California at Santa Cruz, 1156 High Street,
       Santa Cruz, CA 95064}

%%%%%%%%%%%%%%%
% Abstract
%%%%%%%%%%%%%%%

\begin{abstract}
We present several different statistical methods to determine the
transverse velocity vector of M31. The underlying assumptions are that
the M31 satellites on average follow the motion of M31 through space,
and that the galaxies in the outer parts of the Local Group on average
follow the motion of the Local Group barycenter through space. We
apply the methods to the line-of-sight velocities of 17 M31
satellites, to the proper motions of the 2 satellites M33 and IC 10,
and to the line-of-sight velocities of 5 galaxies near the Local Group
turn-around radius, respectively. This yields 4 independent but
mutually consistent determinations of the heliocentric M31 transverse
velocities in the West and North directions, with weighted averages
$\langle v_W \rangle = -78 \pm 41 \kms$ and $\langle v_N \rangle = -38
\pm 34 \kms$. The uncertainties correspond to proper motions of $\sim
10 \uasyr$, which is unlikely to be within reach of direct
observational verification within the next decade. The Galactocentric
tangential velocity of M31 is $42 \kms$, with $1\sigma$ confidence
interval $V_{\rm tan} \leq 56 \kms$. The implied M31--Milky Way orbit
is bound if the total Local Group mass $M$ exceeds
$1.72^{+0.26}_{-0.25} \times 10^{12} \Msun$. If the orbit is indeed
bound, then the timing argument combined with the known age of the
Universe implies that $M = 5.58^{+0.85}_{-0.72} \times 10^{12}
\Msun$. This is on the high end of the allowed mass range suggested by
cosmologically motivated models for the individual structure and
dynamics of M31 and the Milky Way, respectively. It is therefore
possible that the timing mass is an overestimate of the true mass,
especially if one takes into account recent results from the Millennium
Simulation that show that there is also a theoretical uncertainty of
41\% (Gaussian dispersion) in timing mass estimates. The M31
transverse velocity implies that M33 is in a tightly bound orbit
around M31. This may have led to some tidal deformation of M33. It
will be worthwhile to search for observational evidence of this.
\end{abstract}

%%%%%%%%%%%%%%%
% Keywords
%%%%%%%%%%%%%%%


%%%%%%%%%%%%%%%
% Beginning of main text
%%%%%%%%%%%%%%%

\section{Introduction}
\label{s:intro}

The Local Group is dominated by two spiral galaxies, M31 and the Milky
Way. These galaxies have comparable properties, with M31 generally
believed to be slightly more massive (e.g., Klypin, Zhao \& Somerville
2002). The next most luminous galaxies, M33 and the Large Magellanic
Cloud, are some 10 times fainter (e.g., van den Bergh 2000). The
dynamics and future of the Local Group are therefore determined
primarily by the relative velocity of M31 with respect to the Milky
Way. Unfortunately, this velocity is poorly known. The line-of-sight
velocity of M31 can be measured extremely accurately using the Doppler
shift of a large variety of tracers. However, even after a century of
careful attempts (starting with, e.g., Barnard 1917) still no useful
proper motion measurement exists to constrain the transverse
velocity. This limits our ability to answer several fundamental
questions. For example, do M31 and the Milky Way indeed form a bound
system, as is usually assumed (e.g., van den Bergh 1971)? What is the
exact mass of the Local Group implied by the so-called timing argument
(e.g., Kahn \& Woltjer 1959; Kroeker \& Carlberg 1991; Lynden-Bell
1999)? What is the expected future evolution of the M31--Milky Way
system (e.g., Cox \& Loeb 2007)? And how has the structure of M33 been
influenced by possible interaction with M31 (e.g., Loeb
\etal 2005; hereafter L05)?

In the present paper we show that it is possible to obtain a
statistical determination of the transverse velocity of M31 using the
observed velocities of its satellites. The analysis assumes that on
average the satellites follow the motion of M31 through space, with
some velocity dispersion. The ensemble of line-of-sight velocities,
and the individual proper motions available for selected galaxies,
then yield independent estimates of the M31 velocity. We also revisit
the method previously explored by Einasto \& Lynden-Bell (1982), which
is based on the assumption that the galaxies in the outer parts of the
Local Group follow the motion of the Local Group barycenter through
space. We apply the different methods to the currently available data
for the relevant Local Group galaxies, and we combine the results to
obtain an accurate determination of the M31 transverse velocity. We
then use this determination to address the aforementioned questions.

The structure of the paper is as follows. Section~\ref{s:vtrans}
discusses the constraints on the M31 transverse velocity, based
on: the line-of-sight velocities of an ensemble of 17 M31 satellites
(Section~\ref{ss:los}); the recent high accuracy proper motion
determinations of the M31 satellites M33 and IC 10 by Brunthaler \etal
(2005, 2007) from VLBI observations of water masers
(Section~\ref{ss:pm}); and the line-of-sight velocities of an ensemble
of galaxies in the outer parts of the Local group
(Section~\ref{ss:outerlos}). The different constraints are compared
and combined in Section~\ref{ss:combine}. Section~\ref{s:allorbits}
discusses the implications of the inferred M31 velocity for the
relative orbit of M31 and the Milky Way (Section~\ref{ss:orbits}) and
for the relative orbit of M31 and M33 (Section~\ref{ss:triorbits}).
Section~\ref{s:disc} discusses how the results for the M31 transverse
velocity (Section~\ref{ss:vtan}), the total mass of the Local group as
implied by the timing argument (Section~\ref{ss:mass}), and the Local
Group turn-around radius (Section~\ref{ss:turn}) compare with
theoretical predictions and other observational
studies. Section~\ref{s:conc} presents the conclusions of our work.

In the analysis below we use several coordinate systems. Observational
systems have their three principal axes aligned with the
line-of-sight, West, and North directions, respectively, for a given
position on the sky. We also use a Galactocentric coordinate system
centered on the Milky Way and a barycentric coordinate system centered
on the Local Group barycenter. The analysis requires transformations
between the positions and velocities in these systems. Many of the
necessary notations and derivations can be found in van der Marel
\etal (2002; which presents a study of the kinematics of the Large
Magellanic Cloud), and are given here without further
reference. Heliocentric velocities are generally denoted with a vector
${\vec v}$, Galactocentric velocities with a vector ${\vec V}$, and
Local-Group barycentric velocities with a vector ${\vec
U}$. Velocities and proper motions in the observational systems are
generally heliocentric (i.e., not corrected for the reflex motion of
the Sun), unless stated otherwise.

%%% TABLE 1 %%%

\begin{deluxetable}{llrrrrl}
\tabletypesize{\small}
\tablecaption{M31 Satellite Galaxy Sample\label{t:satel}} 
\tablehead{
\colhead{Name} & \colhead{Type} & \colhead{$\rho$} & \colhead{$\Phi$} & \colhead{$v_{\rm los}$} \\
\colhead{} & \colhead{} & \colhead{deg} & \colhead{deg} & \colhead{$\kms$} \\
\colhead{(1)} & \colhead{(2)} & \colhead{(3)} & \colhead{(4)} & \colhead{(5)}
}
\startdata
M31           & Sb I-II   &  0.00 &$\ldots$ & -301 $\pm$ 1 \\
\hline
M32           & E2        &  0.40 & -179.10 & -205 $\pm$ 3 \\
NGC 205       & dSph      &  0.60 &  -46.61 & -244 $\pm$ 3 \\
And IX        & dSph      &  2.69 &   43.32 & -211 $\pm$ 3 \\
And I         & dSph      &  3.27 &  169.84 & -380 $\pm$ 2 \\
And III       & dSph      &  4.97 & -163.13 & -355 $\pm$ 10\\
And X         & dSph      &  5.62 &   48.96 & -164 $\pm$ 3 \\
% And XII       & dSph      &  6.96 &  171.94 & -556 $\pm$ 5 \\
NGC 185       & dSph/dE3  &  7.09 &   -5.07 & -202 $\pm$ 7 \\
NGC 147       & dSph/dE5  &  7.43 &  -12.29 & -193 $\pm$ 3 \\
And V         & dSph      &  8.03 &   35.31 & -403 $\pm$ 4 \\
And II        & dSph      & 10.31 &  136.82 & -188 $\pm$ 3 \\
% And XIV       & dSph      & 11.71 &  170.49 & -478 $\pm$ 5 \\
M33           & Sc II-III & 14.78 &  131.78 & -180 $\pm$ 1 \\
And VII/Cas   & dSph      & 16.17 &  -47.93 & -307 $\pm$ 2 \\
IC 10         & dIrr      & 18.37 &   -9.11 & -344 $\pm$ 5 \\
And VI/Peg    & dSph      & 19.76 & -143.64 & -354 $\pm$ 3 \\
LGS 3/Pisces  & dIrr/dSph & 19.89 &  165.44 & -286 $\pm$ 4 \\
Pegasus       & dIrr/dSph & 31.01 & -143.37 & -182 $\pm$ 2 \\
IC 1613       & dIrr V    & 39.44 &  171.30 & -232 $\pm$ 5 \\
\enddata
\tablecomments{\small The sample of M31 satellites used in the 
modeling of Section~\ref{ss:los}. M31 itself is listed on the first
line for comparison. Column~(1) lists the galaxy name and column~(2)
its type. Columns~(3) and~(4) define the position on the sky: $\rho$
is the angular distance from M31 and $\Phi$ is the position angle with
respect to M31 measured from North over East. These angles were
calculated from the sky positions (RA,DEC) as in van der Marel
\etal (2002). The satellites in the table are sorted by their 
value of $\rho$. Column~(5) lists the heliocentric line-of-sight
velocity and its error.  Velocities and sky positions (RA,DEC) for
most satellites were obtained from the compilation of Evans \etal
(2000), except for the satellites And~IX and~X which had not yet been
discovered in 2000. For those we used the velocity measurements and
(RA,DEC) given by Chapman \etal (2005) and Kalirai
\etal (2007), respectively.}
\end{deluxetable}

%%%%%%%%%%

\section{The M31 Transverse Velocity}
\label{s:vtrans}

\subsection{Constraints from Line-of-Sight Velocities of M31 Satellites}
\label{ss:los}

The velocity vector of an M31 satellite galaxy can be written as the
sum of the M31 velocity vector, and a peculiar velocity
\begin{equation}
\label{vvec}
  {\vec v}_{\rm sat} = {\vec v}_{\rm M31} + {\vec v}_{\rm pec} .
\end{equation}
We assume that any one-dimensional component of ${\vec v}_{\rm pec}$
is a random Gaussian deviate with dispersion $\sigma$. More formally,
this is true if the velocity distribution of the satellites is both
isotropic and isothermal. The first assumption finds some support from
studies of galaxy satellite systems (e.g., Kochanek 1996) and clusters
of galaxies (e.g., van der Marel \etal 2000). The second assumption
is reasonable in view of the fact that the gravitational potentials of
dark halos are approximately logarithmic. Either way, in the present
analysis there is only a very weak dependence on the accuracy of these
assumptions (by contrast to studies of the mass distribution of M31;
e.g., Evans \etal 2000).

The velocity vector of M31 can be described by quantities $v_{\rm
sys}$, $v_t$ and $\Theta_t$, where $v_{\rm sys}$ is the line-of-sight
velocity, $v_t$ is the transverse velocity, and $\Theta_t$ is the
position angle of the transverse motion on the sky. The velocities in
the directions of West and North are
\begin{equation}
\label{vWNdef}
  v_W \equiv v_t \cos (\Theta_t + 90^{\circ}) , \qquad 
  v_N \equiv v_t \sin (\Theta_t + 90^{\circ}) .
\end{equation}
The position of a satellite on the sky can be described by the angles
$(\rho,\Phi)$, where $\rho$ is the angular distance from M31 and
$\Phi$ is the position angle with respect to M31 measured from North
over East (as defined in van der Marel \& Cioni 2001; van der Marel
\etal 2002). Because satellites are not located on the same position on
the sky as M31, the vector ${\vec v}_{\rm M31}$ has a different
decomposition in line-of-sight and transverse components than it does
for M31. More specifically, the line-of-sight velocity of a satellite
is
\begin{equation}
\label{vsatlos}
  v_{\rm sat,los} = 
      v_{\rm sys} \cos \rho +
      v_t \sin \rho \cos (\Phi -\Theta_t) + 
      v_{\rm pec,los} .
\end{equation}
The factor $\cos \rho$ in the first term indicates that only a
fraction of $v_{\rm sys}$ is seen along the line-of-sight. The second
term is an apparent solid body rotation component in the $v_{\rm
sat,los}$ velocity field on the sky, with amplitude $v_t \sin \rho$
and kinematic major axis along position angle $\Theta_t$. The last
term merely adds a scatter $\sigma$ on top of the velocity field
defined by the first two terms.

It follows from equation~(\ref{vsatlos}) that the transverse velocity
of M31 affects the line-of-sight velocities of its
satellites. Therefore, a study of the line-of-sight velocities of the
satellites can constrain the M31 transverse velocity. To this end we
compiled the sample in Table~\ref{t:satel}, which consists of
M31 satellites with known velocities. The sample is based on that used
by Evans \etal (2000), but with addition of the more recently
discovered Andromeda dwarf satellites And~IX and~X. The satellites And
XI and XIII (Martin \etal 2006, who also reports the finding of an
unusually distant globular cluster) and And XV and XVI (Ibata \etal
2007) are not included because no line-of-sight velocity measurements
have yet been reported for them. We fitted equation~(\ref{vsatlos}) to
these data by determining the values of $(v_{\rm sys},v_W,v_N)$ that
minimize the scatter in $v_{\rm pec,los}$. The resulting scatter is
the dispersion $\sigma$ of the satellite population. After the
best-fitting model was identified, we calculated error bars on the
fitted parameters using Monte-Carlo simulations. Many different pseudo
data sets were created with the same satellites at the same positions,
but with velocities drawn from the best-fit model, with $v_{\rm
pec,los}$ drawn as a random Gaussian deviate with dispersion $\sigma$.
The pseudo data sets were then analyzed similarly as the real data
set. The dispersions in the inferred model parameters are a measure of
their formal 1-$\sigma$ error bars.

\begin{figure}[t]
\epsfxsize=0.8\hsize
\centerline{\epsfbox{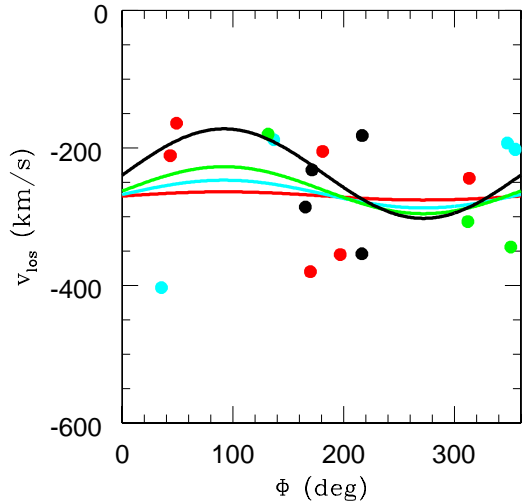}}
\figcaption{Comparison of heliocentric line-of-sight velocity data of M31
satellites and the predictions of equation~(\ref{vsatlos}), for the
best-fit values of the heliocentric M31 transverse velocity $(v_W,v_N)
= (-136,-5) \kms$. The curves show the predictions as function of
position angle $\Phi$, for angular distances from M31 of $\rho =
3^{\circ}$ (red), $\rho = 9^{\circ}$ (cyan), $\rho = 15^{\circ}$
(green), $\rho = 29^{\circ}$ (black). Data points are from
Table~\ref{t:satel}. They are color coded based on bins in angular
distance $\rho$, from 0--6$^\circ$ (red), 6--12$^\circ$ (cyan),
12--18$^\circ$ (green), and 18--40$^\circ$ (black). Predictions with
zero transverse velocity (amplitude zero for the sinusoidal
variations) are also statistically consistent with the data. However,
a large transverse velocity for M31 of hundreds of $\kms$ (corresponding
to sinusoidal variations of much larger amplitude than show here) are
ruled out.\label{f:vlos}}
\end{figure}

The modeling procedure yields $v_{\rm sys} = -270 \pm 19 \kms$, $v_W =
-136 \pm 148 \kms$, $v_N = -5 \pm 75 \kms$, and $\sigma = 76 \pm 13
\kms$ (see also Table~\ref{t:Andvel} below). The data and
representative predictions of the best-fitting model are shown in
Figure~\ref{f:vlos}. Color-coding indicates the distance from M31. The
inferred transverse velocity corresponds to sinusoidal variations that
are less than the observed scatter in the data. No sinusoidal
variation with angle is discernible in the data by eye; indeed, the
case of zero transverse velocity (amplitude zero for the sinusoidal
variations) is also statistically consistent with the data. However,
this is not a null result. Large transverse velocities of hundreds of
$\kms$ would have induced large sinusoidal variations that are not
seen in the data. Such transverse velocities are therefore ruled out.
We note that we could also have kept $v_{\rm sys}$ fixed in the fit at
the observed value of $v_{\rm sys} = -301 \pm 1 \kms$ (Courteau \& van
den Bergh 1999). We verified that such a fit yields verify similar
results, namely $v_W = -123 \pm 159 \kms$ and $v_N = -33 \pm 79 \kms$,
which is well within the uncertainties quoted above.

Our analysis does not assume that the satellites are necessarily bound
to M31. For M33, orbit calculations do suggest that it probably is
bound (see figure 3 of Loeb \etal 2005), but this has not been
established with confidence for most of the other satellites. We
decided not to include the recently discovered dSph galaxies And XII
and And XIV in our sample. They lie at similar position angles, $\Phi
= 171.94^{\circ}$ and $170.49^{\circ}$, respectively, and at distances
$\rho = 6.96^{\circ}$ and $11.71^{\circ}$, that are not atypical for
the rest of our sample.  However, their velocities of $-556 \pm 5$ and
$-478 \pm 5 \kms$, respectively, are $3.3\sigma$ and $2.3\sigma$ away
from the observed velocity of M31. It has been suggested that they
form a system that is falling into M31 for the first time (Chapman
\etal 2007; Majewski \etal 2007), which would not make them useful
additions to our analysis. As a test we did repeat our analysis with
these galaxies included in the sample. This yielded $v_{\rm sys} =
-296 \pm 27 \kms$, $v_W = -85 \pm 200 \kms$, $v_N = 41 \pm
108 \kms$, and $\sigma = 106 \pm 18 \kms$. This is consistent
with the result for our main sample to within the uncertainties.  This
illustrates that the results are fairly robust against the inclusion
or removal of individual galaxies. By contrast, modeling of the parent
galaxy {\it mass} based on satellites can be quite sensitive to
assumptions about the bound state of individual galaxies (e.g.,
Kochanek 1996). 

\subsection{Constraints from Proper Motions of M31 Satellites}
\label{ss:pm}

Water masers can be observed at high spatial resolution with VLBI
techniques. This makes them a valuable tool for proper motion studies
of Local Group galaxies. Brunthaler \etal (2005, 2007) recently
determined the proper motions for two galaxies in the M31 group,
namely M33 and IC 10. Unfortunately, no water masers have yet been
found in M31 and, at present detection limits, may be none should be
expected (Brunthaler \etal 2006). So the direct application of
this technique to M31 may not be possible in the foreseeable
future. However, the measurements for M33 and IC 10 can be used to
constrain the transverse velocity of M31
indirectly. Equation~(\ref{vvec}) implies that the unknown velocity
vector of M31 can be estimated from the known velocity vector of a
satellite as ${\vec v}_{M31} = {\vec v}_{\rm sat} - {\vec v}_{\rm
pec}$. Since the peculiar velocity is unknown, it acts as a Gaussian
uncertainty of size $\sigma$ in each velocity component.

In analogy with equation~(\ref{vsatlos}), one can write for the
transverse velocity components of the satellite
\begin{eqnarray}
\label{vsatprop}
  v_{\rm sat,2} & = & 
      - v_{\rm sys} \sin \rho +
        v_t \cos \rho \cos (\Phi -\Theta_t) + v_{\rm pec,2} , \nonumber \\    
  v_{\rm sat,3} & = & 
      - v_t           \sin (\Phi -\Theta_t) + v_{\rm pec,3} .
\end{eqnarray}
Here the unit vectors 2 and 3 on the plane of the sky are related to
the directions of West and North, all at the position of the
satellite, according to a rotation
\begin{equation}
\label{propmone}
  \left ( \begin{array}{c} v_{{\rm sat}, W} \\ v_{{\rm sat}, N} 
  \end{array} \right )
     =
  \left ( \begin{array}{cc}
     - \sin \Gamma & - \cos \Gamma \\
       \cos \Gamma & - \sin \Gamma \\
  \end{array} \right )
     \>
  \left ( \begin{array}{c} v_{{\rm sat},2} \\ v_{{\rm sat},3}
  \end{array} \right ) , 
\end{equation}
where the rotation angle $\Gamma$ is determined by
\begin{eqnarray}
\label{Gammadef}
   \cos \Gamma & = & [\sin \delta \cos \delta_0 \cos (\alpha-\alpha_0) -
                      \cos \delta \sin \delta_0] \> 
                      / \sin \rho , \nonumber \\    
   \sin \Gamma & = & [\cos \delta_0 \sin (\alpha-\alpha_0)] \> / \sin \rho .
\end{eqnarray}
Here $(\alpha,\delta)$ are the RA and DEC of the satellite, and
$(\alpha_0,\delta_0)$ are the RA and DEC of M31 (i.e., the position
with $\rho = 0$). 

Given values of $(v_{\rm sat,los}, v_{{\rm sat},W}, v_{{\rm sat},N})$,
the equations~(\ref{vWNdef})--(\ref{Gammadef}) uniquely constrain the
three unknown components $(v_{\rm sys}, v_W, v_N)$ of the M31 velocity
vector. We solve these equations for each of the two satellites M33
and IC 10. We take $v_{\rm sat,los}$ from Table~\ref{t:satel}. To
obtain $v_{\rm sat,W}$ and $v_{\rm sat,N}$ we write each velocity
component as $v = (0.0047404 D \mu_{\rm obs}) - \delta v_{\rm rot}$,
where $\mu_{\rm obs}$ is the observed proper motion in $\uasyr$, $D$
is the satellite distance in kpc, and $\delta v_{\rm rot}$ is a
correction for the internal rotation of the galaxy under study. We
take $D = 794 \pm 23 \kpc$ for M33 (McConnachie \etal 2004) and $D =
660 \pm 65 \kpc$ for IC 10 (Evans \etal 2000). The other quantities
follow from Brunthaler et al.~(2005, 2007): $(\mu_{\rm obs,W},\mu_{\rm
obs,N}) = (-4.7 \pm 3.2, -14.1 \pm 6.4) \uasyr$ for M33; $(\mu_{\rm
obs,W},\mu_{\rm obs,N}) = (-6.0 \pm 5.0, 23.0 \pm 5.0) \uasyr$ for
IC10; $(\delta v_{\rm rot,W}, \delta v_{\rm rot,N}) = (70 \pm 23
, -81 \pm 23) \kms$ for M33; and $(\delta v_{\rm rot,W}, \delta
v_{\rm rot,N}) = (-25 \pm 19 , 9 \pm 19) \kms$ for IC 10. We
do not include the Brunthaler \etal corrections for the reflex motion
of the Sun, since we deal with that issue separately in
Section~\ref{ss:orbits}.  We add Gaussian random deviates in our
calculations to reflect the uncertainties. We take each component of
${\vec v}_{\rm pec}$ to be a Gaussian random deviate with dispersion
$\sigma = 76 \kms$, as determined in Section~\ref{ss:los}.  For each
combination of $(v_{\rm sat,los}, v_{\rm sat,W}, v_{\rm sat,N})$ we
solve the equations to obtain $(v_{\rm sys}, v_W, v_N)$ and we repeat
this in Monte-Carlo fashion. We adopt the average and dispersion of
the results as our final estimate for the M31 velocity vector and its
error. Using M33, we obtain the following estimates for M31: $v_{\rm
sys} = -183 \pm 76 \kms$, $v_W = -48 \pm 80 \kms$, $v_N = 71
\pm 84 \kms$. Using IC 10, we obtain the following estimates for
M31: $v_{\rm sys} = -346 \pm 76 \kms$, $v_W = -16 \pm 80
\kms$, $v_N = -47 \pm 81 \kms$ (see also Table~\ref{t:Andvel}
below).

%%% TABLE 2 %%%

\begin{deluxetable*}{llrrlrr}
\tabletypesize{\small}
\tablecaption{Outer Local Group Galaxy Sample\label{t:outer}} 
\tablehead{
\colhead{Name} & \colhead{Type} & \colhead{RA} & \colhead{DEC} & \colhead{$v_{\rm los}$} & \colhead{$D$} & \colhead{$D_{\rm bary}$} \\
\colhead{} & \colhead{} & \colhead{deg} & \colhead{deg} & \colhead{$\kms$} & \colhead{kpc} & \colhead{kpc} \\
\colhead{(1)} & \colhead{(2)} & \colhead{(3)} & \colhead{(4)} & 
\colhead{(5)} & \colhead{(6)} & \colhead{(7)} \\
}
\startdata
WLM             & dIrr IV-V &   0.49234 & -15.46093 & -122 $\pm$  2 &  945 $\pm$  40 &  802 \\
Aquarius/DD0210 & dIrr/dSph & 311.71585 & -12.84792 & -141 $\pm$  2 &  950 $\pm$  50 &  940 \\
Leo A           & dIrr V    & 149.86025 &  30.74639 &   24 $\pm$ \ldots &  800 $\pm$  40 &  953 \\
Tucana          & dSph      & 340.45667 & -64.41944 &  130 $\pm$ \ldots &  870 $\pm$  60 & 1068 \\
Sag DIG         & dIrr V    & 292.49573 & -17.67815 &  -79 $\pm$  1 & 1060 $\pm$ 100 & 1152 \\
\enddata
\tablecomments{\small The sample of outer Local Group galaxies 
used in the modeling of Section~\ref{ss:outerlos}. Column~(1) lists the
galaxy name and column~(2) its type. Columns~(3) and~(4) give the
position on the sky. Column~(5) lists the heliocentric line-of-sight
velocity and, where available, its error. Velocities and sky positions
(RA,DEC) were obtained from the NASA Extragalactic Database (NED).
Column~(6) lists the heliocentric distance and its error, from Grebel
\etal (2003). Column~(7) lists the Local Group barycentric distance 
$D_{\rm bary}$, calculated assuming the M31 distance listed in
Section~\ref{ss:orbits} and a mass fraction $f_{\rm M31} = 0.53$. The
galaxies in the table are sorted by their value of $D_{\rm bary}$.}
\end{deluxetable*}

%%%%%%%%%%

\subsection{Constraints from Line-of-Sight Velocities of Outer Local Group 
Galaxies}
\label{ss:outerlos}

The Local Group contains not only the virialized subgroups of galaxies
surrounding the Milky Way and M31, but also a number of unattached
galaxies that populate the outer regions of the Local Group. On
average, these galaxies are expected to follow the motion of the Local
Group barycenter through space. Their heliocentric line-of-sight
velocity vectors, averaged in a three-dimensional sense, therefore
statistically equal the reflex motion of the Sun with respect to
the Local Group barycenter. Since the velocity of the barycenter is
itself determined by the relative velocity of M31 with respect to the
Milky Way, this yields a determination of the latter.  Variations of
this method have been applied in the past by, e.g., Yahil, Tammann,
\& Sandage (1977), Lynden-Bell \& Lin (1977) and Einasto \&
Lynden-Bell (1982). It is now worthwhile to revisit this method, since
available information on the membership and distances of Local Group
galaxies has evolved significantly in the past decades.

We adopt a Cartesian coordinate system $(X,Y,Z)$, with the origin at
the Galactic Center, the $Z$-axis pointing towards the Galactic North
Pole, the $X$-axis pointing in the direction from the sun to the
Galactic Center, and the $Y$-axis pointing in the direction of the
sun's Galactic Rotation. For the set of outer Local Group galaxies
$i=1,\ldots,N$ we calculate the unit vector ${\vec r}_i$ in the
direction of each galaxy. If the velocity vector of the Sun with
respect to the Local Group barycenter is ${\vec U}_{\odot}$, then one
has
\begin{equation}
\label{matrix}
  \sum_i (v_{{\rm los},i} + [{\vec U}_{\odot} \cdot {\vec r}_i]) {\vec r}_i 
     = 0 ,
\end{equation}
where $v_{{\rm los},i}$ is the heliocentric line-of-sight velocity for
each galaxy. This can be written as a $3 \times 3$ matrix equation for
the components of the vector ${\vec U}_{\odot}$. The best-fit values
and their formal errors are easily obtained using standard techniques
(Einasto \& Lynden-Bell 1982). Once this solution is obtained one can
calculate the velocity of the Milky Way with respect to the Local
Group Barycenter, ${\vec U}_{\rm MW} = {\vec U}_{\odot} - {\vec
V}_{\odot}$. Here ${\vec V}_{\odot}$ is the velocity vector of the Sun
in the Galactocentric rest frame. For the circular velocity of the
Local Standard of Rest (LSR) we use the standard IAU value $V_0 = 220
\kms$ (Kerr \& Lynden-Bell 1986), to which we assign an uncertainty
$10 \kms$ (none of our results depend sensitively on this
quantity). For the residual velocity of the Sun with respect to the
LSR we adopt the values of Dehnen \& Binney (1998). If we assume that
all of the mass of the Local Group resides in the Milky Way and
M31, then the barycenter is simply the mass-weighted average of their
position vectors. This implies in the Galactocentric rest frame that
the velocity of M31 is ${\vec V}_{\rm M31} = - {\vec U}_{\rm MW} /
f_{\rm M31}$, where $f_{\rm M31} \equiv M_{\rm M31} / (M_{\rm M31} +
M_{\rm MW})$. The heliocentric velocity of M31 is therefore ${\vec
v}_{\rm M31} = {\vec V}_{\rm M31} - {\vec V}_{\odot}$. After
substitution of the previous equations this yields
\begin{equation}
\label{outersol}
  {\vec v}_{\rm M31} = - {\vec U}_{\odot}/ f_{\rm M31} + 
                         {\vec V}_{\odot} [ (1/f_{\rm M31}) - 1] . 
\end{equation}
This heliocentric vector can be decomposed into components along the
line-of-sight and in the West and North directions following the
methodology of van der Marel \etal (2002).

Application of equation~(\ref{outersol}) requires that we assume a
value for $f_{\rm M31}$. Einasto \& Lynden-Bell (1982) used the
Tully-Fisher relation to constrain this quantity, and built this
constraint directly into their matrix solution for ${\vec
U}_{\odot}$. However, the mass ratio of M31 and the Milky Way isn't
actually all that well known observationally, and different arguments
for estimating it have yielded different results. Our aim here is to
constrain ${\vec v}_{\rm M31}$ from observational data, while
imprinting a minimum amount of theoretical prejudice into the
result. So we simply assume that $f_{\rm M31}$ is homogeneously
distributed between $0.39$ and $0.67$ (i.e., $M_{\rm M31} / M_{\rm MW}
= 0.8$--$2.0$). This encompasses most of the values that have been
quoted in the literature (see discussion in Section~\ref{ss:mass}). We
then solve equation~(\ref{outersol}) in Monte-Carlo fashion, while
simultaneously adding in the random errors in ${\vec U}_{\odot}$ and
${\vec V}_{\odot}$. This yields both the best fit result for ${\vec
v}_{\rm M31}$ and its statistical uncertainties.

We base our analysis on the sample of Local Group dwarf galaxies
(there are no giant galaxies in the outer parts of the Local Group)
compiled by Grebel, Gallagher, \& Harbeck (2003). From their table~1
we removed all galaxies listed as being (potentially) part of the
Milky Way or M31 subgroups. We obtained the heliocentric line-of-sight
velocities of the galaxies from the NASA Extragalactic Database
(NED). We removed the Cetus dwarf from the sample, since it has no
line-of-sight velocity available. We restricted our primary sample
to galaxies with a Local Group barycentric distance less than $\sim
1.2 \Mpc$. The resulting sample consists of 5 galaxies, which are
listed in Table~\ref{t:outer}. With this sample the analysis yields
for M31: $v_{\rm sys} = -405 \pm 114 \kms$, $v_W = -126 \pm 63
\kms$, $v_N = -89 \pm 50 \kms$ (see also Table~\ref{t:Andvel} below). 
The fit yields an estimate $\sigma = 22 \kms$ for the
one-dimensional dispersion of the galaxies around the space motion of
the Local-Group barycenter. The small value of this dispersion is due
to the fact that the sample galaxies reside at an average Local Group
barycentric distance of $0.98 \Mpc$, which is consistent with the
Local Group turn-around radius (e.g., Karachentsev
\etal 2002), where the velocity with respect to the Local Group
barycenter is zero by definition.

When comparing the analyses in Section~\ref{ss:los} and the present
section, there is an important difference in the expected galaxy
velocities. The velocities of satellites around M31 are virialized, so
the expectation value of a satellite velocity with respect to M31 is
zero, independent of where the satellite is located on the sky. By
contrast, the motions of the outer Local-Group galaxies around the
Local Group barycenter are not virialized. Therefore, the expectation
value of a galaxy velocity with respect to the Local Group barycenter
is zero only if the galaxy is near the turn-around radius.  If this is
not the case, then equation~(\ref{matrix}) is valid only if the
galaxies are distributed homogeneously around the Local Group.  In
reality, the distribution is both non-homogeneous and the number of
galaxies is small. Therefore, addition to the sample of galaxies
beyond the turn-around radius is expected to add both bias and
shot-noise to the estimate of the M31 velocity. Moreover, galaxies
significantly outside the turn-around radius do not necessarily need
to follow the Local Group barycenter motion.

The sample in Table~\ref{t:outer} consists of a rather small number of
galaxies. So despite the aforementioned disadvantages, we did study
the effect of adding more Local Group galaxies at larger barycentric
distances. In particular, we tried to add the only other 6 Local Group
galaxies within $2 \Mpc$ (namely: NGC 3109, Antlia, Sextans A and B,
IC 5152 and KKR 25; Grebel \etal 2003) that are not believed to be
associated with any other nearby structures. These reside at
barycentric distances of 1.6--$1.9
\Mpc$. Analysis of the combined sample of 11 galaxies yields for M31:
$v_{\rm sys} = -608 \pm 154 \kms$, $v_W = -82 \pm 138 \kms$,
$v_N = -46 \pm 82 \kms$.  The fit yields an estimate $\sigma =
50 \kms$ for the one-dimensional dispersion of the galaxies around
the space motion of the Local-Group barycenter. The large deviation of
$v_{\rm sys}$ from the observed value as well as the increased
$\sigma$ and formal errors support our assertion that adding these
distant galaxies decreases the quality of the results.  Nonetheless,
the results for $v_W$ and $v_N$ are consistent with those
inferred from the smaller sample, within the errors. So we conclude
that the results for the M31 transverse motion are quite robust, and
not very sensitive to the composition of the sample.  This is further
supported by the fact that our results are consistent within the
errors with the preferred solutions obtained by Einasto \& Lynden-Bell
(1982), despite their use of a sample that is only partially
overlapping with ours.

%%% TABLE 3 %%%

\begin{deluxetable}{llll}
\tabletypesize{\small}
\tablecaption{M31 Heliocentric Velocity Estimates\label{t:Andvel}} 
\tablehead{
\colhead{Method} & \colhead{$v_{\rm sys}$} & \colhead{$v_W$} & \colhead{$v_N$} \\
\colhead{} & \colhead{$\kms$} & \colhead{$\kms$} & \colhead{$\kms$} \\
\colhead{(1)} & \colhead{(2)} & \colhead{(3)} & \colhead{(4)}
}
\startdata
M31 Satels.    & -270 $\pm$  19 & -136 $\pm$ 148 &   -5 $\pm$  75 \\
M33 PM         & -183 $\pm$  76 &  -48 $\pm$  80 &   71 $\pm$  84 \\
IC 10 PM       & -346 $\pm$  76 &  -16 $\pm$  80 &  -47 $\pm$  81 \\
Outer LG Gals. & -405 $\pm$ 114 & -126 $\pm$  63 &  -89 $\pm$  50 \\
\hline
Weighted Av.   & -273 $\pm$  18 &  -78 $\pm$  41 &  -38 $\pm$ 34
\enddata
\tablecomments{\small Estimates of the heliocentric velocity 
of M31 estimated using different methods, as indicated in column~(1).
The method based on the M31 satellite ensemble line-of-sight
velocities is described in Section~\ref{ss:los}, and that based on the
observed proper motions (PMs) of M33 and IC 10 is described in
Section~\ref{ss:pm}, and that based on the line-of-sight velocities of
the satellites in outer regions of the Local Group is described in
Section~\ref{ss:outerlos}. Column~(2) lists the estimated M31
systemtic line-of-sight velocities. Columns~(3) and~(4) lists the
estimated M31 transverse velocities in the West and North directions,
respectively. The bottom line of the table lists the weighted average
of the results from the different methods.}
\end{deluxetable}

%%%%%%%%%%

\subsection{Comparison and Combination of Constraints}
\label{ss:combine}

\begin{figure}[t]
\epsfxsize=0.8\hsize
\centerline{\epsfbox{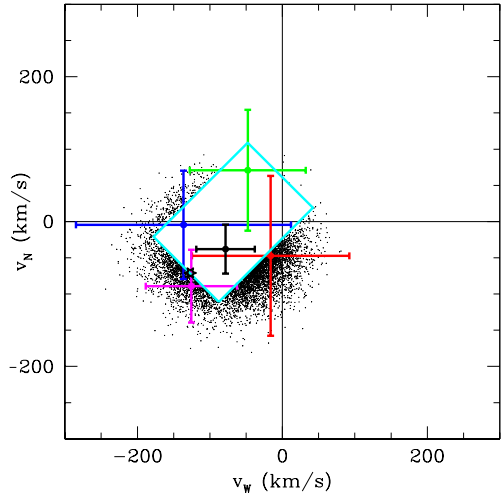}}
\figcaption{Estimates of the M31 heliocentric transverse velocity
in the West and North directions. Data points with error bars are from
Table~\ref{t:Andvel}, based on the M31 satellite ensemble
line-of-sight velocities (blue), the proper motion of M33 (green), the
proper motion of IC 10 (red), the line-of-sight velocities of outer
Local group satellites (magenta), or the weighted average of these
results (black). The starred symbol indicates the transverse velocity
that corresponds to a radial orbit for M31 with respect to the Milky
Way. The cyan rectangle approximates the region that is ruled out in
all of the theoretical models explored by Loeb \etal (2005), because
the resulting relative orbit of M31 and M33 would have produced more
disruption of the M33 disk than is observed. The small dots are the
18\% of 30,000 samplings from the error ellipse belonging to the black
data point that are consistent with the theoretical
constraint.\label{f:vwvn}}
\end{figure}

The $v_W$ and $v_N$ for M31 inferred from the different methods and
listed in Table~\ref{t:Andvel} are shown in Figure~\ref{f:vwvn} as
colored data points with error bars. The weighted averages of all four
of the independent estimates are $\langle v_W \rangle = -78 \pm 41
\kms$ and $\langle v_N \rangle = -38 \pm 34 \kms$. This is shown in
the figure as a black data point with error bars.

The $\chi^2$ that measures the residuals between the individual
measurements in Table~\ref{t:Andvel} and the weighted averages is
$\chi^2 = 8.1$ for $N = 9$ degrees of freedom (12 measurements minus 3
parameters). Therefore, the results for the $v_W$ and $v_N$ from the
different methods are consistent within the errors. Among other
things, this implies that there is no evidence that the dispersion of
the peculiar velocities of M31 satellites in the transverse direction,
which enters into the analysis of Section~\ref{ss:pm}, is larger than
the value $\sigma = 76 \kms$ derived from line-of-sight velocities in
Section~\ref{ss:los}. This is consistent with the assumption that was
made about the isotropy of the peculiar velocities.

The inferred weighted average of the three independent systemic
velocity estimates is $\langle v_{\rm sys} \rangle = -273 \pm 18
\kms$. This differs at the $1.6\sigma$ level from the observed M31
velocity $v_{\rm sys} = -301 \pm 1 \kms$ (Courteau \& van den Bergh
1999), but the agreement in an absolute sense is better than the
formal uncertainties in $(\langle v_W \rangle, \langle v_N
\rangle)$. So this comparison provides no reason to mistrust our
assumptions that the M31 satellites (Sections~\ref{ss:los}
and~\ref{ss:pm}) and the outer Local Group galaxies
(Section~\ref{ss:outerlos}) move on average through space with the
same velocity as M31 and the Local Group barycenter, respectively (see
Lynden-Bell 1999 for an earlier discussion of this).

The analyses in the previous sections make assumptions about the
velocity distributions of the satellites, but not about their spatial
distributions. These spatial distributions are known to be
inhomogeneous. In the case of the M31 subgroup, there are more
satellites on the near side of M31 (i.e., between M31 and the Milky
Way) than on the far side (e.g., McConnachie \& Irwin 2006). Moreover,
recent studies have suggested that (some of) the satellites are
concentrated near a plane surrounding M31 (e.g., Koch \& Grebel
2006). These facts by themselves do not affect our analysis at all, as
long as the velocity distributions remain isotropic. However, the
analysis {\it would} be affected if the satellite ensemble possessed a
mean rotation. This is possible in some of the scenarios that have
been suggested for a possible disk-like distribution of satellites
(but not necessarily the favored scenarios; see, e.g., Metz, Kroupa,
\& Jerjen 2007). In the method of Section~\ref{ss:los} we fit an
apparent solid-body rotation field (eq.~[\ref{vsatlos}]) to the
observed satellite velocities. Any intrinsic rotation would therefore
bias the inferred transverse velocity. However, if this had been the
case then we might have expected the M31 transverse velocity results
of this method to be inconsistent with the results from the other
methods that we have used, and in particular the method of
Section~\ref{ss:outerlos} based on outer Local Group galaxies. The
fact that the results from the different methods are actually
statistically consistent therefore suggests that our results are not
affected by potential rotation of the M31 satellite system.

In the comparisons of the different estimates listed in
Table~\ref{t:Andvel} it should be noted that they are not completely
independent, because the line-of-sight velocities of M33 and IC10
enter not only in the analysis of Section~\ref{ss:pm}, but also in
that of Section~\ref{ss:los}. However, this is not a very important
issue. We verified that if M33 and IC10 are removed from the
line-of-sight velocity analysis in Section~\ref{ss:los}, then the
changes in the results are well within the uncertainties.

\section{Orbits}
\label{s:allorbits}

\subsection{M31--Milky Way Orbit}
\label{ss:orbits}

To calculate the velocity of M31 in the Galactocentric rest frame we
adopt the same Cartesian coordinate system $(X,Y,Z)$ as in
Section~\ref{ss:outerlos}. We adopt a distance $D = 770 \pm 40 \kpc$ for
M31 (Holland 1998; Joshi \etal 2003; Walker 2003; Brown \etal 2004;
McConnachie \etal 2005; Ribas \etal 2005). The position of M31 is then
${\vec r} = (-379.2,612.7,-283.1) \kpc$. To calculate the reflex motion
of the Sun at the position of M31 we use the same solar velocity as in
Section~\ref{ss:outerlos} and we use the standard IAU value $R_0 = 8.5
\kpc$ for the distance of the Sun from the Galactic Center (Kerr \&
Lynden-Bell 1986), to which we assign an uncertainty of $0.5 \kpc$
(none of our results depend sensitively on this quantity). The
velocity of the Sun then projects to $(v_{\rm sys}, v_W, v_N)_{\odot}
= (172, 128, 71) \kms$ at the position of M31. Since one
observes the reflex of this, these values must be {\it added} to the
observed M31 velocities to obtain its velocity in the Galactocentric
rest frame. The velocity vector corresponding to the observed velocity
component $v_{\rm sys}$ given in Table~\ref{t:satel} and the inferred
$(\langle v_W \rangle, \langle v_N \rangle)$ given in
Table~\ref{t:Andvel} is then ${\vec V}_{\rm obs} = (97 \pm 35 ,
-67 \pm 26 , 80 \pm 32) \kms$. The errors (which are
correlated between the different components) were obtained by
propagation of the errors in the individual position and velocity
quantities (including those for the Sun) using a Monte-Carlo scheme.

\begin{figure*}[t]
\epsfxsize=0.8\hsize
\centerline{\epsfbox{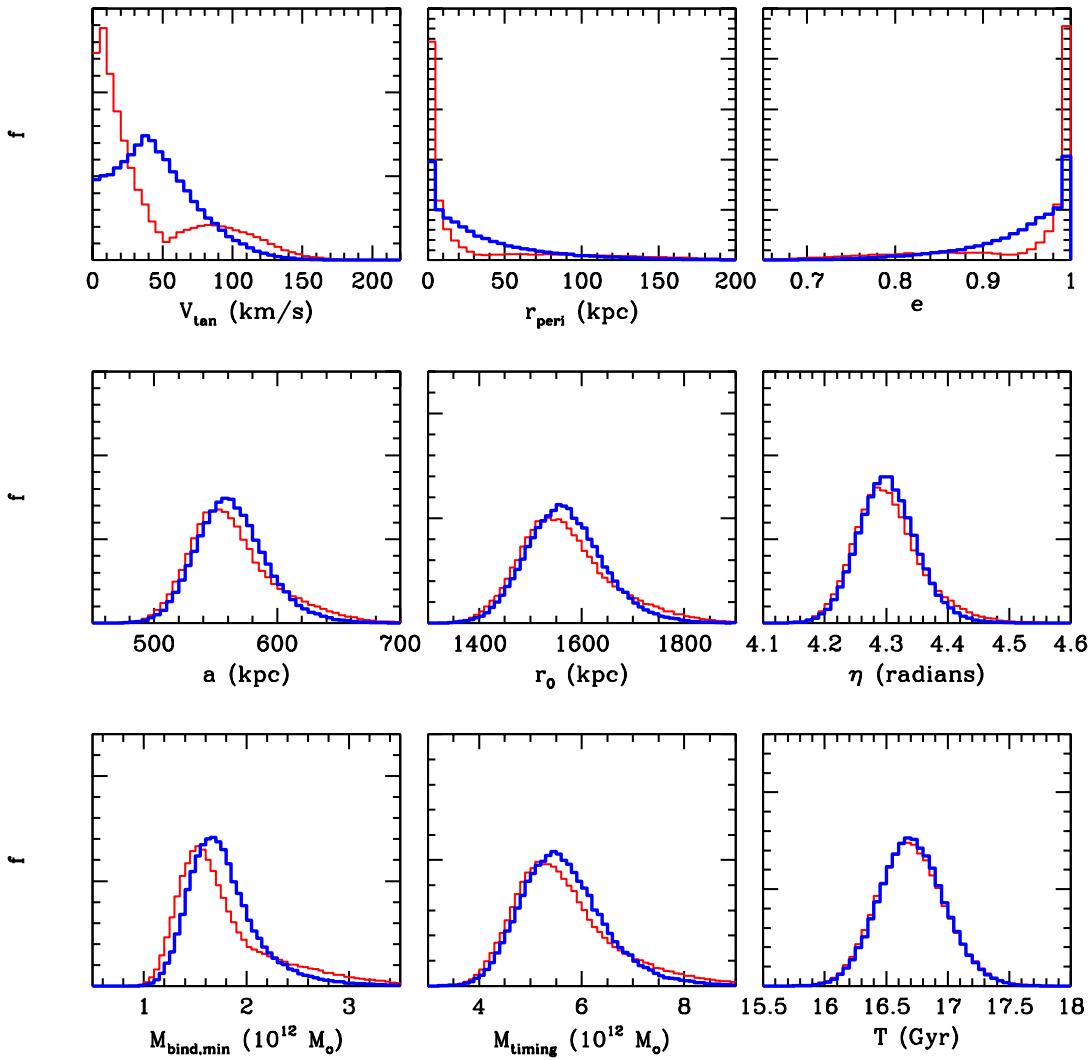}}
\figcaption{Probability histograms of: M31 
Galactocentric tangential velocity $V_{\rm tan}$; minimum total mass
$M_{\rm bind,min}$ of M31 and the Milky Way if the galaxies are bound;
and assuming a bound orbit and the timing argument, the M31-Milky Way
orbital pericenter distance $r_{\rm peri}$, orbital eccentricity $e$,
orbital semi-major axis length $a$, orbital period $T$, current
eccentric anomaly $\eta$, total mass $M_{\rm timing}$, and implied
turnaround radius $r_0$ for test-particles on radial orbits around the
M31-Milky Way system. The probability distributions were obtain using
Monte-Carlo simulations as described in the text. The vertical scales
are arbitrary. The blue curves take into account all observational
uncertainties in the distances and velocities of both M31 and the Sun
in the Galactocentric rest frame, as well as the observational
uncertainties in the age of the Universe.  The red curves also enforce
the theoretical exclusion zone of Loeb \etal 2005 (cyan region in
Figure~\ref{f:vwvn}), within which more tidal deformation of M33 would
have been expected than is observed.\label{f:hist}}
\end{figure*}

If the transverse velocity of M31 in the Galactocentric rest frame,
$V_{\rm tan}$, equals zero, then M31 moves straight towards the Milky
Way on a purely radial (head-on collision) orbit. This orbit has
$(v_W,v_N)_{\rm rad} = (-127, -71) \kms$ (this is approximately
the reflex of the velocity of the Sun quoted above, because the lines
from the Sun to M31 and from the Galactic Center to M31 are almost
parallel). The radial orbit is indicated as a starred symbol in
Figure~\ref{f:vwvn}. The velocity ${\vec V}_{\rm obs}$ calculated in
the previous paragraph has tangential and radial components $V_{\rm
tan,obs} = 59 \kms$ and $V_{\rm rad,obs} = -130 \kms$. The total
velocity is $|{\vec V}_{\rm obs}| = 142 \kms$.

The value $V_{\rm rad,obs}$ is an unbiased estimator of the true
radial velocity $V_{\rm rad}$. The associated uncertainty can be
calculated in straightforward fashion using the previously described
Monte-Carlo scheme. This yields $V_{\rm rad} = -130 \pm 8 \kms$.
The uncertainty is due mostly to the uncertainty in our knowledge of
the circular velocity of the LSR in the Galactic plane. The values of
$V_{\rm tan,obs}$ and $|{\vec V}_{\rm obs}|$ are more difficult to
interpret, because they are not unbiased estimators of the true
velocities $V_{\rm tan}$ and $|{\vec V}|$. This is because the area
coverage of $V_{\rm tan}$ values in the $(v_W,v_N)$ plane scales as $2
\pi V_{\rm tan} d V_{\rm tan}$, which produces a bias in the sense
that any measurement error in $(v_W,v_N)$ tends to yield overestimates
of $V_{\rm tan}$ and $|{\vec V}|$. To quantify and correct these
biases we used Bayes' theorem, which yields the identity
\begin{equation}
  P(V_{\rm tan} | V_{\rm tan, obs}) \propto 
    P(V_{\rm tan}) P (V_{\rm tan, obs} | V_{\rm tan})  
\end{equation}
We are interested in the quantity on the left-hand side, which is the
probability distribution of $V_{\rm tan}$, given our measurement.  The
quantity $P(V_{\rm tan})$ on the right-hand side is the Bayesian prior
probability of $V_{\rm tan}$, which we assume to be flat (i.e.,
homogeneous). The quantity $P (V_{\rm tan, obs} | V_{\rm tan})$ is the
probability of measuring a value $V_{\rm tan, obs}$ if the actual
value is $V_{\rm tan}$. This latter distribution is easily calculated
using Monte-Carlo drawings, because the measurement uncertainties are
known. Once the distributions $P (V_{\rm tan, obs} | V_{\rm tan})$
have been pre-calculated for all $V_{\rm tan}$, it is straightforward
to obtain a Monte-Carlo sampling of the probability distribution
$P(V_{\rm tan} | V_{\rm tan, obs})$. To this end one draws a random
deviate $V_{\rm tan}$, and then accepts this value with probability $P
(V_{\rm tan, obs} | V_{\rm tan})$. The top left panel of
Figure~\ref{f:hist} shows the probability distribution $P(V_{\rm tan}
| V_{\rm tan, obs})$ thus obtained. The median $V_{\rm tan} = 42
\kms$. The fact that this is smaller than $V_{\rm tan, obs} = 59 \kms$
quantifies the aforementioned bias. The $1\sigma$ confidence interval
is $V_{\rm tan} \leq 56 \kms$. So the radial orbit is consistent
with the data at this confidence level. These results are discussed in
the context of previous model predictions in
Section~\ref{ss:vtan}. When combined with the value for $V_{\rm rad}$,
the $1\sigma$ confidence interval around the median for the total
velocity is $|{\vec V}| = 138^{+14}_{-11} \kms$.

To get insight into the relative orbit of M31 and the Milky Way we
assume that they can be approximated as point masses of mass $M_{\rm
M31}$ and $M_{\rm MW}$, respectively. In the center-of-mass frame,
their orbit then has energy $E = {1\over2} \mu |{\vec V}|^2 - G \mu M
/ \vert {\vec r} \vert$ and angular momentum $L = \mu \vert {\vec r}
\times {\vec V} \vert$, where the total mass is $M = M_{\rm M31} +
M_{\rm MW}$ and the reduced mass is $\mu = M_{\rm M31} M_{\rm MW} /
M$. The galaxies are bound to each other ($E<0$) if $M \geq M_{\rm
bind, min} = \vert {\vec r} \vert \> |{\vec V}|^2 / 2 G$. The orbit 
of the separation vector ${\vec r}$ is then a Kepler
ellipse with eccentricity $e^2 = 1 + (2 E L^2 / G^2 M^2 \mu^3)$ and
semi-major axis length $a = L^2 / [G M \mu^2 (1-e^2)]$. The pericenter
separation is $r_{\rm peri} = a / (1-e)$.  The period is $T = 2 \pi
(a^3/GM)^{1/2}$. The orbit can be parameterized using the eccentric
anomaly $\eta$ as (e.g., Kibble 1985; Binney \& Tremaine 1987)
\begin{eqnarray}
\label{Keplerpos}
  r & = & a(1-e \cos\eta) , \nonumber \\
  t & = & (a^3/GM)^{1/2} (\eta-e\sin\eta) .
\end{eqnarray}
In this parameterization $\eta = 0$ corresponds to the pericenter
passage at $t=0$, while $\eta = 2\pi$ corresponds to the next
pericenter passage at $t=T$. The radial and tangential velocities can
be similarly parameterized as (e.g., Kochanek 1996)
\begin{eqnarray}
\label{Keplervel}
  V_{\rm rad} & = & (GM/a)^{1/2} (e \sin\eta) / (1 - e \cos\eta) , \nonumber \\
  V_{\rm tan} & = & (GM/a)^{1/2} (1 - e^2)^{1/2} / (1 - e \cos\eta) .
\end{eqnarray}

It is generally assumed that the Local Group is a bound system that,
due to its overdensity, decoupled from the Hubble expansion at fairly
high redshift. Since then the orbital evolution of M31 and the Milky
Way has been governed by Newtonian dynamics. Given this scenario and
the orbital description given by equation~(\ref{Keplerpos}), the Big
Bang must have corresponded to a previous pericenter passage, which we
can take to be $t=0$. The current time $t$ then corresponds to the age
of the Universe, which has been tightly constrained using data from
the Wilkinson Microwave Anisotropy Probe to be $t =
13.73^{+0.16}_{-0.15} \Gyr$ (Spergel \etal 2007). With measurements of
the current Galactocentric distance $r$ and velocities $V_{\rm rad}$
and $V_{\rm tan}$, the equations~(\ref{Keplerpos})
and~(\ref{Keplervel}) can be solved for the quantities $M$, $a$, $e$
and $\eta$. The quantity $\eta$ must be in the interval $[\pi,2\pi]$
(so that M31 and the Milky Way are falling towards each other for the
first time), since unplausibly high masses $M$ are otherwise required.

This methodology for modeling the Local Group is commonly called the
``timing-argument''. It has been widely applied and discussed in the
literature (e.g., Kahn \& Woltjer 1959; Lynden-Bell 1981, 1999;
Einasto \& Lynden-Bell 1982; Raychaudhury \& Lynden-Bell 1989; Kroeker
\& Carlberg 1991; Kochanek 1996; Loeb \etal 2005), mostly to obtain joint
estimates for the age of the Universe and the mass of the Local Group,
often assuming a purely radial orbit. We can now do a more accurate
analysis, both because we have an observational estimate of $V_{\rm
tan}$, and because the age of the Universe can be assumed to be
well-known from independent data. We have used this methodology on our
results. Monte-Carlo error propagation was performed as described
above, to obtain full probability distributions of $M$, $a$, $e$,
$\eta$, $T$, and $r_{\rm peri}$. The results are shown as histograms
in Figure~\ref{f:hist}.  The inferred $1\sigma$ confidence intervals
around the median are: $M = 5.58^{+0.85}_{-0.72} \times 10^{12}
\Msun$; $a = 561^{+29}_{-26} \kpc$; $\eta = 4.301^{+0.047}_{-0.045}$ 
radians; $r_{\rm peri} = 23 \kpc$, with $1\sigma$ confidence interval
$r_{\rm peri} \leq 40.9 \kpc$; $T = 16.70^{+0.27}_{-0.26} \Gyr$; and
$e = 0.959$, with $1\sigma$ confidence interval $(1-e) \leq
0.072$. The implications of this result for $M$ are discussed in
Section~\ref{ss:mass}. The uncertainties in the listed quantities are
due primarily to the uncertainties in $V_{\rm tan}$ and the M31
distance $D$. The uncertainties in these quantities contribute more or
less equally to the uncertainties in $M$, $a$, $T$, and $\eta$. The
uncertainties in the distance have little effect on $e$ and $r_{\rm
peri}$. All quantities vary monotonically with $V_{\rm tan}$ and
$D$. Larger values of $V_{\rm tan}$ yield larger values of $M$, $a$,
$\eta$, $T$, and $r_{\rm peri}$, and smaller values of $e$. Larger
values of $D$ yield larger values of $M$, $a$, $e$, $T$, and $r_{\rm
peri}$, and smaller values of $\eta$. Figure~\ref{f:hist} also shows
the probability distribution of $M_{\rm bind, min}$, the minimum mass
required for a bound orbit. The inferred $1\sigma$ confidence
intervals around its median is $M_{\rm bind, min} =
1.72^{+0.26}_{-0.25} \times 10^{12} \Msun$.

The anticipated collision between M31 and the Milky Way will happen at
the next orbital pericenter, which in the Kepler model is at $T-t
\approx 3.0 \pm 0.3 \Gyr$ from now. The orbital pericenter distance
is much smaller than the sizes of the galaxies' dark halos. If the
galaxies are assumed to have a logarithmic potential that reproduces
the observed rotation curve amplitudes, then their radial extents
$r_t$ are $235 (M_{\rm M31}/3.4 \times 10^{12} \Msun) \kpc$ and $207
(M_{\rm MW}/2.3 \times 10^{12} \Msun) \kpc$, respectively (Loeb \etal
2005).  The dark halos will intersect once the orbital separation
becomes smaller than the sums of these sizes.  This will have two
consequences. First, the orbits will deviate from Kepler ellipses in
the sense that the orbit will have less curvature and larger $r_{\rm
peri}$ than indicated by the previously derived Kepler orbit. Second,
there will be dynamical friction, which will tend to increase the
curvature and will tend to decrease $r_{\rm peri}$. More complicated
calculations are necessary to properly calculate the orbit once the
dark halos of the galaxies start to overlap and to study the
properties of the resulting merger. Such calculations were recently
presented by Cox \& Loeb (2007). However, they adopted an orbit with
$V_{\rm tan} = 132 \kms$, $r_{\rm peri} = 450 \kpc$, and $e =
0.494$. Comparison to the top panel of Figure~\ref{f:hist} shows that
this orbit is not consistent with our measurement of the M31
transverse velocity.

\subsection{M31--M33 Orbit}
\label{ss:triorbits}

Loeb \etal (2005; hereafter L05) recently derived a theoretical
constraint on the transverse motion of M31 from the fact that M33 is
relatively undisturbed. This appears to rule out orbits where M33 had
a previous close interaction with M31. The exact shape of the region
in $(v_W,v_N)$ space thus ruled out (being defined by L05 as: more
than 20\% of the M33 stars would have been stripped) has a complex
shape and depends somewhat on the modeling assumptions, but it can be
approximated by the solid cyan rectangle in Figure~\ref{f:vwvn}. This
is an approximation to figure~2c of L05 (which quantified the
transverse motion using Galactocentric rest-frame velocities
$V_{\alpha {\rm cos} \delta} = -v_W - 128 \kms$ and $V_{\delta} =
v_N + 71 \kms$). In some of their models (see their figures~2a,b)
even a somewhat larger region is ruled out.

The overall shape of the excluded region in $(v_W,v_N)$ space can be
understood with fairly simple calculations and arguments. In doing so,
we ignore the observational uncertainties in the quantities of
interest. This is sufficient for the scope of the present discussion,
but should be included for more quantitative understanding of the
M31-M33 orbit. Based on the data from Table~\ref{t:satel} and the M33
and M31 distances from Sections~\ref{ss:pm} and~\ref{ss:orbits},
respectively, the distance between M33 and M31 is $203 \kpc$. The
center of the excluded region in Figure~\ref{f:vwvn} is at $(v_W,v_N)
\approx (-68,-11) \kms$. This defines the three-dimensional velocity
vector of M31, while for M33 that vector is known from Brunthaler
\etal (2005). The radial and tangential velocity components of the
separation vector are then $V_{\rm rad} = -71 \kms$ and $V_{\rm tan} =
126 \kms$. The Kepler orbit of the separation vector can be calculated
similarly as in Section~\ref{ss:orbits}. In doing so, we assume that
$M_{\rm M31} = 3.0 \times 10^{12}$, which is based on $M = 5.58 \times
10^{12} \Msun$ (Section~\ref{ss:orbits}) and $f_{\rm M31} \approx
0.54$ (compare Section~\ref{ss:outerlos}). We assume that the
M33-to-M31 mass ratio is equal to the value $\sim (118/250)^4$
suggested by the Tully-Fisher relation and the galaxy's circular
velocities (Corbelli \& Schneider 1997; Klypin \etal 2002). The orbit
then has $a = 120 \kpc$, $r_{\rm peri} = 27 \kpc$, $T = 2.2 \Gyr$, and
$e = 0.77$. The value of $a$ is close to the minimum that this
quantity can attain (i.e., maximum binding energy) as a function of
$(v_W,v_N)$, which is $a_{\rm min} = 113 \kpc$. Given the value of
$a$, M33 moves inside of the M31 dark halo for most of its orbit. Our
assumption that all of the M31 mass resides at its center therefore
overestimates the curvature of the orbit. The listed pericenter
distance should thus be interpreted as a lower bound on the actual
value. More detailed calculations, as in L05, are required to get a
proper estimate of this quantity.

As one moves from $(v_W,v_N) \approx (-68,-11) \kms$ along the
diagonal that runs from the bottom left to the top right in
Figure~\ref{f:vwvn}, the Kepler orbit value of $r_{\rm peri}$ doesn't
change much.  However, $a$ increases, which means that the orbits
become more eccentric and less bound (and ultimately unbound). The
galaxies therefore spend less time in close vicinity of each other, and
the relative velocity at pericenter increases. By contrast, as one
moves from $(v_W,v_N) \approx (-68,-11) \kms$ along the diagonal that
runs from the top left to the bottom right in Figure~\ref{f:vwvn}, the
Kepler orbit value of $e$ doesn't change much. But again, $a$
increases, which means that the orbits have larger pericenter
separations and become less bound (and ultimately unbound). The energy
dissipated during an encounter scales as $dE \propto 1/(r^4 V^2)$ in
the impulsive tidal approximation (e.g., Binney \& Tremaine 1987,
eq.~[7-55]), where $r$ is the impact parameter. Therefore, as one
moves away along either diagonal in Figure~\ref{f:vwvn}, the structure
of M33 will be less perturbed by the encounter. The excluded region in
Figure~\ref{f:vwvn} can therefore be understood to lowest order as the
region where the orbital integral over $dE$ exceeds some threshold,
with $r$ and $V$ calculated on the basis of the Kepler orbit.

Interestingly, the observationally implied transverse velocity of M31
from Table~\ref{t:Andvel} falls right in the region for which
considerable tidal deformation of M33 would have been expected. This
M31 velocity yields a Kepler orbit for the M31--M33 separation vector
with $a = 127 \kpc$, $r_{\rm peri} = 30 \kpc$, $T = 2.4 \Gyr$, and $e
= 0.76$. However, the uncertainties in the observationally
implied velocities cannot be ignored. Upon performing Monte-Carlo
sampling we find that as much as 18\% of samplings from the error
ellipse fall outside the cyan rectangle in Figure~\ref{f:vwvn}, and
therefore do not violate the M33 tidal stripping argument. These 18\%
are shown as small black dots for a total sample of 30,000
drawings. The black dots can be viewed as a visual representation of
the probability distribution of M31's transverse velocity obtained by
taking into account not only the observational constraints derived
here, but also the theoretical M33 stripping argument of L05. Since
most of the dots fall close to the cyan rectangle, it is likely that
there has been some tidal deformation of M33 by M31 (although not
sufficient to pass L05's threshold for being considered
excluded). This suggests that it will be worthwhile to perform deep
searches for tidal tails and structures in the outer regions of M33,
similar to those that have already been performed for M31 (e.g.,
Ferguson \etal 2002).

Our results for the relative M31--M33 orbit involve a small amount of
circular reasoning, since we have assumed a priori that the residual
space motion of M33 with respect M31 has the same dispersion ($\sigma
= 76 \kms$ per coordinate, as derived from line-of-sight velocities in
Section~\ref{ss:los}) as do the other M31 satellites. To avoid this
circular reasoning one could ignore the result labeled ``M33 PM'' in
Table~\ref{t:Andvel} (green cross in Figure~\ref{f:vwvn}).  The
weighted average of the remaining M31 transverse motion estimates then
becomes $\langle v_W \rangle = -89 \pm 47 \kms$ and $\langle v_N
\rangle = -60 \pm 37 \kms$. This differs by only $24 \kms$ from what
we have used so far (last line of Table~\ref{t:Andvel}), and this
difference is well within the uncertainties. This modified result
still falls well inside the region excluded by L05. Therefore, our
conclusions about the M31--M33 orbit are not influenced by the fact
that we have used the M33 space velocity as one of the estimators of
the M31 space velocity.

L05 only considered models with $M_{\rm M31} = 2.6$--$3.4 \times
10^{12} \Msun$. As is discussed in Section~\ref{ss:mass} below, it is
possible that the M31 mass is actually lower than this. For example,
Klypin \etal (2002) advocate $M_{\rm M31} = 1.6 \times 10^{12}
\Msun$. It follows both from the simple arguments of
Section~\ref{ss:triorbits} and from the detailed calculations of L05
(their figure~2) that the amount of past tidal deformation of M33 is
smaller for smaller values of $M_{\rm M31}$. This would reduce the
area of the excluded rectangular region in Figure~\ref{f:vwvn}, and
would reduce the concern that our observational estimate of the M31
transverse velocity falls in the region of parameter space that was
disfavored by L05. Also, the models of L05 (and hence the cyan region
in Figure~\ref{f:vwvn}) do not account explicitly for the
observational uncertainties in the assumed distances of M31 and
M33. If the actual distances differ at the 1 or 2$\sigma$ level from
the canonical values, then this could affect the location of the
excluded region in transverse-velocity space.

It is straightforward to include the theoretical constraint of L05
into the M31-Milky Way timing argument calculations of
Section~\ref{ss:orbits}. To address this, we applied the same
Monte-Carlo scheme as in that section, but now with rejection of all
Monte-Carlo drawn velocities with $(v_W,v_N)$ combinations in the
region excluded by L05. This yields the probability histograms shown
in red in Figure~\ref{f:hist}. The distributions of $V_{\rm tan}$, $e$
and $r_{\rm peri}$ become rather non-Gaussian, as can be easily
understood from the distribution of points in
Figure~\ref{f:vwvn}. However, the other distributions remain close to
Gaussian, and are not very different from those obtained without using
the LO5 M33 stripping argument. The same is true for the inferred
$1\sigma$ confidence intervals around the median values. For example,
for the Local Group mass we obtain $M = 5.50^{+1.14}_{-0.76} \times
10^{12} \Msun$, which is similar to the result of
Section~\ref{ss:orbits}.

\section{Discussion}
\label{s:disc}

\subsection{M31 Tangential Velocity}
\label{ss:vtan}

Many previous studies of the M31--Milky Way system have assumed that
their orbit can be approximated to be radial ($V_{\rm tan} \approx
0$). This simplifies analyses based on the timing argument and, in the
absence of a reliable $V_{\rm tan}$ measurement, was a reasonable
guess based on simple cosmological arguments. In the absence of mutual
gravitational interactions, peculiar velocities with respect to the
Hubble flow decrease with time as $(1+z)$. In the M31--Milky Way
system there is mutual gravitational attraction along the galaxy
separation vector. This changes the radial velocity $V_{\rm rad}$ from
positive (receding) at high $z$ to negative (approaching) at the
present time, as quantified by the timing argument. However, the
angular momentum is conserved in a two-body system without external
perturbations. Therefore, a significant present-day tangential motion
in such as system implies an unrealistically high peculiar velocity at
high redshift.
 
The situation is more complicated when the possibility of angular
momentum exchange is taken into account. Tidal torques can lead to
exchange between the spin angular momentum of the galaxies and
their orbital angular momentum. This process may have contributed both
to the observed spins of M31 and the Milky-Way, and to the tangential
velocity component in their orbital motion (Gott \& Thuan 1978; Dunn
\& Laflamme 1993). More importantly, tidal torques exerted by the
galaxies outside of the Local Group also induce a tangential velocity
component in the M31--Milky Way system (Raychaudhury \& Lynden-Bell
1989). A useful approach to study this effect, and more generally, the
orbits of all the galaxies in the nearby Universe, is based on the
principle of least action (Peebles 1989).  This assumes that nearby
galaxies arrived at their present configuration through gravitational
interactions from a nearly homogeneous high-redshift state with
negligible peculiar velocities. Allowed solutions are those that
minimize the relevant Hamiltonian action integral. This method can be
used to fit observed galaxy velocities starting from their observed
positions (Peebles \etal 1989; Peebles 1994), to fit observed galaxy
distances starting from their observed velocities (Shaya, Peebles, \&
Tully 1995; Schmoldt \& Saha 1998), or to fit observed galaxy
velocities and distances simultaneously (Peebles \etal 2001, hereafter
P01; Pasetto \& Chiosi 2007). The general prediction from the
theoretical work that includes tidal torques is that M31 tangential
velocities of $V_{\rm tan} \lta 200 \kms$ are expected in plausible
models.\footnote{The mass $M = M_{\rm M31} + M_{\rm MW}$ calculated
with the timing argument is a monotonically increasing function of
$V_{\rm tan}$. So if one pre-assumes an upper limit to $M$, then one
also obtains an upper limit to $V_{\rm tan}$. L05 used this approach
to obtain $V_{\rm tan} \lta 120 \kms$ based on the assumption that $M
\leq 5.6 \times 10^{12} \Msun$. However, no physical motivation was
provided for this assumed mass limit. Larger values of $V_{\rm tan}$
are not inconsistent with the timing argument, but they do require
higher masses.} 

Figure~6 of P01 shows the M31 transverse velocity vectors predicted in
30 minimum-action solutions for the nearby Universe. The reason that
there are multiple possible solutions is due to the absence of
observational knowledge of most galaxy proper motions. P01
characterized the M31 velocity using supergalactic angular coordinates
in the Galactocentric rest frame. These are related to the
heliocentric transverse velocity $v_t$ and the position angle
$\Theta_t$ of the transverse motion on the sky, defined as in
equation~(\ref{vWNdef}), according to
\begin{eqnarray}
\label{vSGLdef}
  v_{\rm SGL} & = & - v_t \cos (\Theta_t + 34.60^{\circ}) 
                     - 131 \kms, \nonumber \qquad \\
  v_{\rm SGB} & = & \>\>\>  v_t \sin (\Theta_t + 34.60^{\circ}) 
                     - 64 \kms .
\end{eqnarray}
Our weighted average velocity $(\langle v_W \rangle, \langle v_N
\rangle)$ given in Table~\ref{t:Andvel} corresponds to
$(v_{\rm SGL},v_{\rm SGB}) = (-55 , -21) \kms$.  The $68.3$\%
confidence ellipse around this measurement encloses 8 of the 30 viable
solutions presented by P01. Our measurement is therefore fully
consistent with their theoretical work. The action method is based on
the assumptions that mass follows light and that the galaxy peculiar
velocities are due to their mutual gravitational interactions. Our M31
transverse motion determination therefore provides no reason to doubt
these assumptions (although cosmological N-body simulations suggest
that these assumptions are at best only approximately satisfied;
Martinez-Vaquero, Yepes, \& Hoffman 2007). The mass $M_{\rm M31} +
M_{\rm MW}$ assumed in the P01 models is $5.16
\times 10^{12} \Msun$, which is consistent with the range 
calculated in Figure~\ref{f:hist} based on the timing argument. The
mass was not varied independently in P01, but is within the factor
$\sim 2$ range of masses for which the action method yields plausible
solutions (Peebles \etal 1994). Pasetto \& Chiosi (2007) obtained a
best-fit solution from their action modeling that corresponds to a
heliocentric velocity $(v_W,v_N) = (-142, -41) \kms$. This value
is near the edge of our $68.3$\% confidence ellipse, and is therefore
also consistent with our measurement. The mass $M_{\rm M31} + M_{\rm
MW}$ assumed in their models is $5.36 \times 10^{12} \Msun$, which is
also within the range calculated in Figure~\ref{f:hist} based on the
timing argument.

In summary, our measurement of the transverse velocity of M31 is
consistent with the most recent theoretical models. Moreover, the fact
that 73\% of P01's action-method solutions do {\it not} fall within
our $68.3$\% confidence ellipse suggests that our measurement has
sufficient accuracy to provide meaningful constraints on the allowed
solution space of the Local Group orbits and its history.

\subsection{Local Group Mass}
\label{ss:mass}

In Section~\ref{ss:orbits} we calculated that the minimum $M = M_{\rm
M31} + M_{\rm MW}$ required for a bound orbit is $M_{\rm bind, min} =
1.72^{+0.26}_{-0.25} \times 10^{12} \Msun$ and that for a bound orbit
the timing argument implies that $M = 5.58^{+0.85}_{-0.72}
\times 10^{12} \Msun$. These results are based on point mass dynamics, 
where the point masses are proxys for entire the Milky Way subgroup of
galaxies and the entire M31 subgroup of galaxies (i.e., the parent
galaxy plus its satellites), respectively. Only a negligible fraction
of the light in the Local Group comes from galaxies that are not part
of either of these subgroups (see Table~\ref{t:outer}). Therefore, the
mass $M$ is a proxy for the total mass of the Local Group.

It is of interest to compare our mass estimates to those derived using
independent arguments. Kochanek (1996) modeled the mass distribution
of the Milky Way using equilibrium models for the velocities of Milky
Way satellites, with the circular velocity of the disk and the
escape velocity of stars in the solar neighborhood as additional
constraints. This yielded a fairly well constrained mass of $(0.5 \pm
0.1) \times 10^{12} \Msun$ inside of $50
\kpc$. However, the total Milky Way mass depends on the extent
of the dark halo, which is poorly constrained by any data.  Kochanek's
solutions therefore allowed total masses anywhere from $\sim$ 1--5
$\times 10^{12} \Msun$ (see his figure~7).  Wilkinson \& Evans (1999)
obtained a similarly large range from their models, $M =
1.9^{+3.6}_{-1.7} \times 10^{12} \Msun$, for the same reason. Both of
these studies used rather arbitrarily parameterized density
profiles. Klypin, Zhao \& Somerville (2002; hereafter K02) improved
this situation by restricting the discussion to models with
$\Lambda$CDM cosmologically motivated density profiles and
concentrations. This yielded a much more tightly constrained total
mass. Their favored solution (their model A1) has a total virial mass
of $1.0 \times 10^{12} \Msun$. The highest-mass model that is still
consistent with the observational and theoretical constraints (their
model A4; see also Besla \etal 2007) has a total virial mass of $2.0
\times 10^{12} \Msun$. 

Evans \& Wilkinson (2000) and Evans \etal (2000) modeled the mass
distribution of M31 using equilibrium models for the velocities of M31
satellites and halo tracers. This yielded $M = 0.8^{+1.6}_{-0.5}
\times 10^{12} \Msun$, suggesting that M31 be less massive 
than the Milky Way. However, as for the Milky Way modeling, the
allowed mass range was large due to uncertainties in the dark halo
extent. Again, K02 improved this situation by imposing theoretical
constraints from $\Lambda$CDM modeling. This yielded a favored model
(their model B1) with a total virial mass of $1.6 \times 10^{12}
\Msun$. The ratio $M_{\rm M31} / M_{\rm MW} = 1.6$ for the favored
models in the K02 study (this implies $f_{\rm M31} = 0.62$, as defined
in Section~\ref{ss:outerlos}). This is similar to the value implied by
the ratio of the galaxy's circular velocities and the Tully-Fisher
relation, $\sim (250/220)^4 = 1.67$ (Einasto \& Lynden-Bell 1982).

K02 did not explore what maximum mass might still be consistent with
the observational data for M31. However, to lowest order one may guess
that the situation is similar as for the Milky Way, where the highest
allowed mass is twice the favored mass. The maximum Local Group mass
for the K02 models is therefore $\sim 5.2 \times 10^{12} \Msun$, while
the favored mass is $M = 2.6 \times 10^{12} \Msun$. This assumes that
the derived virial masses of the Milky Way and Andromeda apply to
their entire subgroups of satellites. This is reasonable, given the
uncertainties already inherent in the estimates. The largest
satellites of both subgroups (the Large Magellanic Cloud and M33,
respectively) have $\lta 10$\% of the luminosity of their parent
galaxy, and most other satellites contribute $\lta 1$\%. Both the
favored and the maximum mass of K02 exceed the value $M_{\rm bind,
min} = 1.72^{+0.26}_{-0.25} \times 10^{12} \Msun$ derived in
Section~\ref{ss:orbits}. This suggests that the Milky Way and M31 are
indeed a bound system, as is usually assumed.

We showed that the timing argument for a bound M31--Milky Way system
implies a mass $M = 5.58^{+0.85}_{-0.72} \times 10^{12} \Msun$. This
is consistent with the maximum mass allowed by the K02 models, but not
with the favored mass. Even the tail of the probability distribution
for $M_{\rm timing}$ in Figure~\ref{f:hist} doesn't go as low as $M =
2.6 \times 10^{12} \Msun$. The best agreements are obtained for the
case of nearly radial orbits and a small M31 distance. For example,
$V_{\rm tan} = 0$ and $D = 700 \kpc$ yields $M_{\rm timing} = 4.4
\times 10^{12} \Msun$. But while this $V_{\rm tan}$ is consistent with
our measurements, such a low distance is almost $2\sigma$ away from
the value $D = 770 \pm 40 \kpc$ inferred using a wide range of methods
(Holland 1998; Joshi \etal 2003; Walker 2003; Brown \etal 2004;
McConnachie \etal 2005; Ribas \etal 2005).

The timing argument provides a rather simplistic view of the evolution
of a binary galaxy in a cosmological context. Several authors have
therefore quantitatively studied its accuracy. Models that address the
gravitational influence and torques of structures outside of the Local
Group have generally concluded that these do not significantly bias
the mass estimates obtained from the timing argument (Raychaudhury \&
Lynden-Bell 1989; Peebles \etal 1989, 2001; Pasetto \& Chiosi 2007;
see also Section~\ref{ss:vtan}). However, these models generally treat
individual galaxies as point masses. So this does not address the fact
that the galaxies themselves assemble hierarchically over time.

To include hierarchical assembly it is necessary to perform more
detailed dynamical calculations. Kroeker \& Carlberg (1991) studied
the accuracy of the timing argument by examining binary galaxies found
in an $N$-body simulation of a closed CDM Universe. They found that
when $V_{\rm tan}$ is available, the timing argument provides an
unbiased estimate of the total mass (measured in two spheres, centered
on each galaxy, of radii one half the inter-galaxy separation). But
estimates based on radial orbits yielded masses that were on average
too low by a factor $\sim 1.7$. They also found that the galaxies must
be bound for the timing argument to work, since it overestimates the
actual mass of unbound pairs. The independent mass estimates of the
Milky Way and M31 discussed above suggest that the galaxies are in
fact bound, so this should not be an issue here. Kroeker \& Carlberg
also show that the timing argument may not work well for pairs that
are not isolated. This too should not be much of an issue for the
M31--Milky Way system, given that the Sculptor, M81, and Maffei groups
are all some $3 \Mpc$ away.

Goldberg (2001) reached somewhat different conclusions than Kroeker \&
Carlberg. He modeled binary galaxies using a perturbative least action
approach in which the galaxies are part of a continuous fluid that
collapses over the course of the simulation. The results were
cross-validated against an N-body simulation, with consistent
results. From his calculations he concluded that the timing argument
tends to overestimate the total mass of a pair by a factor of $\sim
2$. The reason for the discrepancy between this result and that of
Kroeker \& Carlberg is unclear. However, Goldberg's calculations are
quite idealized, while Kroeker \& Carlberg's calculations are now
quite dated (both in terms of numerical sophistication and underlying
cosmology). A more modern theoretical investigation of the accuracy of
timing argument was therefore presented recently by Li \& White
(2008).\footnote{This paper came to our attention after the original
submission of our own work.}

Li \& White identified galaxy pairs in the so-called Millennium
Simulation with properties similar to the Milky Way--M31 system. The
high quality and detail of this cosmological simulation provides
excellent statistics, with thousands of pairs available for
study. They found that the {\it radial-orbit} timing mass estimator
provides an unbiased estimate of the total mass of a pair (defined as
the sum of the $M_{200}$ values, where $M_{200}$ is the mass inside
the sphere over which the mean density is 200 times that of the
Universe at large). With their preferred selection cuts, they found
the ratio of the true to the estimated mass to be $A = 0.99 \pm 0.41$,
where we transformed their quoted interquartile range to a traditional
Gaussian dispersion. The simulations do not confirm the finding of
Kroeker \& Carlberg (1991) that the bias in the estimate depends on
$V_{\rm tan}$, but they do indicate that the dispersion in the
estimate increases with $V_{\rm tan}$. Either way, the radial-orbit
analysis of Li \& White is reasonable for the Milky Way--M31 system,
since we have found that a radial orbit is consistent with the
available observational constraints. The calculated ``theoretical
uncertainty'' exceeds the uncertainties calculated here (see
Figure~\ref{f:hist}) that result from propagation of observational
errors. This theoretical uncertainty therefore drives the accuracy
which we the Local Group mass is known from the timing argument. The
preferred Local Group mass of K02 corresponds to a ratio $A=0.47$. This is
$1.2\sigma$ away from timing mass that we have calculated (with
$\sigma$ being the quadrature sum of the theoretical and observational
uncertainties). Therefore, the two results are quite consistent. K02
quote a mass uncertainty of a factor $\sim 2$, which is not too
dissimilar from the timing mass uncertainty calculated by Li \& White
(2008). Therefore, it is not a priori clear which estimate is to be
preferred.

\subsection{Local Group Turn-Around Radius}
\label{ss:turn}
    
An independent mass estimate is obtained from the turn-around radius
$r_0$ of the Local Group, defined as the distance from the Local Group
barycenter at which galaxies have a radial velocity of zero (e.g.,
Lynden-Bell 1981). Using the same Keplerian formalism as in
Section~\ref{ss:orbits}, galaxies on the turn-around surface have $\eta
= \pi$ and $V_{\rm rad} = 0$. Equation~(\ref{Keplerpos}) therefore
gives that
\begin{equation}
\label{turnaround}
  r_0 = [G M (t/\pi)^2]^{1/3} (1+e) ,
\end{equation}
where, as before, $t$ is the age of the Universe and $r_0$ is usually
evaluated for radial orbits ($e=1$). Figure~\ref{f:hist} shows the
probability distribution of the predicted $r_0$ thus obtained, with
$M$ taken from the timing argument. The inferred $1\sigma$ confidence
interval around the median is $r_0 = 1.56^{+0.08}_{-0.07} \Mpc$. 

The $r_0$ predicted from the M31--Milky Way timing argument can be
compared to the observationally inferred value, $r_0 = 0.94 \pm 0.10
\Mpc$ (e.g., Karachentsev \etal 2002). These values are not
consistent. Since $r_0 \propto M^{1/3}$, the observed $r_0$ implies a
Local Group mass of only $M = (1.3 \pm 0.3) \times 10^{12} \Msun$.
Similar results are obtained when this method is applied to other
nearby groups (Karachentsev \etal 2005). This mass is lower than that
implied by the timing argument (as stressed previously by Lynden-Bell
1999) and it is also lower than that implied by almost all studies of
the individual masses of the Milky Way and M31 (see
Section~\ref{ss:mass}). Moreover, comparison of this mass to $M_{\rm
bind, min} = 1.72^{+0.26}_{-0.25} \times 10^{12} \Msun$ from
Section~\ref{ss:orbits} suggests that M31 and the Milky Way then do
not form bound pair. This would make the Local Group little more than
a chance superposition of two spiral galaxies. Such a configuration
has fairly low a priori probability, given the local density of spiral
galaxies (van den Bergh 1971).

The agreement between the observed and predicted turn-around radii
improves if the galaxies defining the turn-around surface are not on
radial orbits. A value $e = 0.2$ in equation~(\ref{turnaround}) would
yield perfect agreement. However, the corresponding orbits are then
nearly circular and this would produce a considerable thickness in the
zero-velocity surface which is inconsistent with observations
(Lynden-Bell 1999). So a more reasonable conclusion may be that mass
estimates based on the turn-around radius are systematically biased
towards low values. Direct evidence for this comes from the action
modeling of, e.g., Peebles \etal (2001). Their models reproduce both
the distances and redshifts of the galaxies in the outer regions of
the Local Group (see Table~\ref{t:outer}), despite the use of a Local
Group mass $M = 5.16 \times 10^{12} \Msun$ that significantly exceeds
the mass implied by equation~(\ref{turnaround}). This suggests that
the isolated collapse picture on which the concept of a turn-around
surface is based may not be suitable for the Local Group. The
underlying reason may be that the dynamical structure near the edges
of the Local Group is significantly influenced by structures in the
mass distribution on larger scales. It should also be noted that the
even in idealized models the zero-velocity surface is not actually
expected to be a sphere, but an elongated ellipsoid (e.g., figure~6 in
Peebles \etal 1989). This is not generally taken into account in
observational determinations of $r_0$.

\section{Conclusions}
\label{s:conc}

We estimated the transverse motion of M31 under the assumption that
the satellites of M31 follow the motion of M31 through space, and
under the assumption that the galaxies in the outer parts of the Local
Group follow the motion of the Local Group barycenter through space.
The first method was applied independently to a sample of 17
satellites with known line-of-sight velocities, and to the 2
satellites M33 and IC 10 with known proper motions. The second method
was applied to a sample of 5 galaxies near the Local Group turn-around
radius (and we showed that inclusion of 6 more distant galaxies does
not significantly affect the results). The results from the different
methods are mutually consistent and successfully recover the known
line-of-sight velocity of M31. The weighted average heliocentric
transverse velocity of M31 from the different methods in the West and
North directions are found to be $\langle v_W \rangle = -78 \pm 41
\kms$ and $\langle v_N \rangle = -38 \pm 34 \kms$.

We used the known line-of-sight velocity with the newly inferred
transverse velocity to determine the radial and tangential
Galactocentric velocity components of M31. We used a Bayesian analysis
to obtain the statistical probability distribution for $V_{\rm tan}$,
properly corrected for observationally induced errors and biases. The
results are that $V_{\rm rad} = -130 \pm 8 \kms$ and that the
probability distribution for $V_{\rm tan}$ has a median value of $42
\kms$ and a $1\sigma$ confidence interval $V_{\rm tan} \leq 56
\kms$. A purely radial orbit is consistent with the data at this
confidence level. These velocities are consistent with the predictions
of the most recent action-method models for the history of the Local
Group, and have small enough errors that they can start to
meaningfully constrain the full solution space allowed by these
models.

We used the M31 velocity vector to constrain the relative M31--Milky
Way orbit. The minimum mass required for a bound orbit is $M_{\rm
bind, min} = 1.72^{+0.26}_{-0.25} \times 10^{12} \Msun$. If the orbit
is indeed bound, then the timing argument combined with the known age
of the Universe implies that $M = 5.58^{+0.85}_{-0.72} \times 10^{12}
\Msun$. This is consistent with mass range suggested by cosmologically
motivated models for the individual structure and dynamics of M31 and
the Milky Way, respectively. However, the timing mass is on the high
end of the allowed range, with the most favored models in the study of
K02 yielding a Local Group mass of only $2.6 \times 10^{12}
\Msun$. This indicates that the M31 and the Milky Way are indeed
bound, but that their total mass could be lower than suggested by the
timing argument. Li \& White (2008) performed a theoretical
calibration of the timing argument and found that while the timing
argument mass is unbiased, it does have a theoretical dispersion of
41\%. Therefore, it is not statistically implausible that for the
specific case of the Milky Way--M31 system the timing argument mass
could be an overestimate by a factor of $\sim 2$. The method of
estimating the Local Group mass based on the size of the turn-around
(zero-velocity) radius yield systematically lower masses than other
methods, and is in fact not consistent with a bound nature for the
M31--Milky Way orbit. This and other arguments suggest that this
method yields incorrect masses that are systematically biased low.

The M31 transverse velocity implies that M33 is in a tightly bound
orbit around M31. The calculations of L05 indicate that for 82\% of
Monte-Carlo samplings from the error ellipse, more tidal deformation
of M33 would have been expected than is observed.  However, this
assumes that $M_{\rm M31} = 2.6$--$3.4 \times 10^{12} \Msun$. The
percentage is lower, and the agreement between observational and
theoretical constraints on the M31 transverse velocity is better, if
the M31 mass is as low as $1.6 \times 10^{12} \Msun$, as advocated by
K02. Nonetheless, our results indicate that some tidal deformation of
M33 could certainly have occurred. So it will be worthwhile to perform
more deep searches for tidal tails and structures in the outer regions
of M33. The estimates of the Local Group timing mass do not change
significantly if the theoretical constraints from the L05 study of M33
structure are strictly enforced.

The transverse velocity inferred for M31 (the weighted average in
Table~\ref{t:Andvel}) corresponds to a proper motion $\mu_W \equiv
-{\dot \alpha} \cos \delta = -21.5 \pm 11.1 \uasyr$ and $\mu_N \equiv
{\dot \delta} = -10.4 \pm 9.3 \uasyr$ at a distance $D = 770 \kpc$. It
will be difficult to obtain an actual proper motion measurement in the
near future that will rival the accuracy of this. The most accurate
Hubble Space Telescope proper motion measurements for any Local Group
galaxy (for the Large and Small Magellanic Cloud; Kallivayalil \etal
2006a,b) have errors of 50--$80 \uasyr$. VLBI measurements of water
masers can provide errors of only 5--$10 \uasyr$ (Brunthaler \etal
2005, 2007). This is actually not much better than what we have
already presented here. Moreover, no water masers are known in M31 to
which this technique can be applied and, at present detection limits,
may be none should be expected (Brunthaler \etal 2006). Therefore, a
high accuracy measurement of the proper motion of M31 will have to
await future generations of astrometric satellites. However, even SIM
or GAIA measurements will at best have uncertainties that are only a
few times smaller than those reported here (see section~12 of Unwin
\etal 2007), and moreover, results are not expected from these
satellites for at least another decade.

It may be possible to improve the results presented here by using a
larger number of tracers of the M31 velocity. If the sample of M31
satellites with water maser proper motions can be increased from 2 to
$\sim 8$, then this would halve the uncertainty in the average result
from the method in Section~\ref{ss:pm}. However, this may not be
possible. Water masers are often associated with regions of star
formation, whereas most M31 satellites are early type dwarf
ellipticals or spheroidals (see Table~\ref{t:satel}).  Even in the
gas-rich satellites the prospects for new water maser discoveries are
not good (Brunthaler \etal 2006). Therefore, the number of satellites
for which water maser proper motions can be obtained is fundamentally
limited. Alternatively, it may be possible to increase the number of
tracers at large distances from M31 for which line-of-sight velocities
are available. A significant increase would be needed (from 17 to
$\sim 68$) to merely halve the uncertainties in our result from
Section~\ref{ss:los}. While new satellites of M31 continue to be
discovered on a regular basis (e.g., Zucker \etal 2007; Ibata \etal
2007), this is unlikely to significantly affect the
statistics. Individual Red Giant Branch stars in the outer halo of M31
may be more promising (e.g., Gilbert \etal 2006), since it is
possible, at least in principle, to obtain line-of-sight velocities
for large numbers of such stars.

%%%%%%%%%%%%%%%
% Acknowledgments
%%%%%%%%%%%%%%%

\acknowledgments 

We thank Nitya Kallivayalil and Gurtina Besla for a careful reading of
this paper, and Anatoly Klypin, Frank van den Bosch, Lars Hernquist,
and T.J. Cox for useful discussions. We thank the anonymous referee
for helpful suggestions that improved the presentation of the paper.
We made use of the NASA/IPAC Extragalactic Database (NED), which is
operated by the Jet Propulsion Laboratory, Caltech.

%%%%%%%%%%%%%%%
% Appendix
%%%%%%%%%%%%%%%

% Uncomment if there are appendices
% \appendix

%%%%%%%%%%%%%%%
% Start references on a new page, unless we are in emulate mode
%%%%%%%%%%%%%%%

%%%%%%%%%%%%%%%
% Reference List
%%%%%%%%%%%%%%%

{}

\end{document}